\documentstyle[11pt,epsfig]{article}
\title{\bf Vacuum spherically symmetric solutions in  $f(T)$ gravity}
\author{K. Atazadeh\thanks{email: atazadeh@azaruniv.ac.ir}\,\ and
Misha Mousavi\thanks{email: mousavi@azaruniv.ac.ir}
\\ {\small
Department of Physics, Azarbaijan Shahid Madani University , Tabriz, 53714-161 Iran}}
\begin{document}
\maketitle

\begin{abstract}
Spherically symmetric static vacuum solutions have been built in $f(T)$ models of gravity theory. We apply some conditions on the metric components; then the new vacuum spherically symmetric solutions are obtained. Also, by extracting metric coefficients we determine the  analytical form of $f(T)$.
\end{abstract}
\vspace{2cm}
\section{Introduction}
It is difficult to obtain explicit solutions of modified gravitational field equations on account of their nonlinear character, for instance the equations of motion in General Relativity (GR) and $f( R)$ theory, respectively, are of order 2 and 4. However, there are a reasonable number of static spherically symmetric exact solutions which have a measure of physical interest, particularly solutions of $f(R)$ gravity theories in vacuum space \cite{vilja,capoz} and for global static sphere with finite radius at which the pressure vanishes, called stellar model \cite{shojai}. Clearly, we are interested in analyzing such a kind of results for vacuum space, in solar system tests; then we can get to know with the exterior gravitational field surrounding some massive spherical object such as a star. We can use the extracted metrics to investigate the physics in the vicinity of a spherical object, in particular the trajectories of freely falling massive particles and photons.

A new group of models which has been added recently to the candidate class of models for explaining the present day acceleration of our universe is known as $f(T)$ gravity theories \cite{rafael2}-\cite{myrza}. The idea of $f(T)$ gravity theory refers to 1928 when Einstein was trying to redefine the unification of gravity and electromagnetism by means of the introduction of a tetrad (vierbein) field together with the idea of absolute parallelism \cite{einstein}. In the teleparallel  gravity (TG) theories the dynamical object is not the metric $g_{\mu\nu}$ but a set of tetrad fields $e_{a}(x^{\mu})$ and rather than the well-known torsionless Levi-Civita connection of GR, a Weitzenb\"{o}k connection is used to define the covariant derivative; torsion plays the role of curvature in TG \cite{Aldrovandi}-\cite{de}. The degree of non-linearity in $f(T)$ field equations is the same as the order of GR.

A crucial point about the $f(T )$ is that it does
not respect local Lorentz symmetry \cite{barrow,barrow3}. From a theoretical
perspective this is a rather undesirable feature and experimentally
there are stringent constraints. A Lorentz--violating
theory is only attractive if the violations are
small enough to avoid detection and it leads to some other
significant achievements. So far, the only pay-off that has
been suggested is that $f(T )$ gravity might provide an
alternative to conventional dark energy in general relativistic
cosmology.

In accordance with the very recent attention to spherically symmetric space--time in $f(T)$ gravity, a large number of vacuum and non-vacuum solutions have been built in this theory \cite{wang,rafael}. In searching for solutions of the field equations of $f(T)$ gravity models, considering vacuum solutions of nonlinear second-order field equations of $f(T)$ gravity theory comes first.
 In \cite{nashed} exact spherically symmetric solutions of $f(T)$ theories by very different methods are studied. The usual way of finding the complete vacuum model with exact solution necessitates us to start with the form of $f(T)$, then replace it in modified field equations to find the metric coefficients.
In this paper we follow a different strategy and construct modified field equations thus rewriting them in such a manner as to make the nonlinear differential two reduced nonlinear differential equations somewhat easy, to obtain a variety of explicit solutions by means of inserting some constrains in the metric coefficients to make considerable simplification. This method  is exactly the same as Tolman strategy which put varied relations for the metric components to solve the reduced Einstein equations in an easy way. As a result we introduce three relations for $X(r)$, $F(r)$ and $A(r)$ as additional constraints on field equations and solve the equations, thereafter we can extract the form of $f(T)$. It should be noted that this it is not guaranteed that all these constraints produce $f(T)$ in a physically analytical form.

This paper is organized as follows: in the section $2$, we consider some basic concepts of $f(T)$ theory, and in the presence of the locally Lorentz violation  we do some calculations on the field equations to convert those into the covariant version according to the approach that has been introduced in \cite{barrow2}. In the section 3, by using the covariant version of the field equations  we discuss different situations for the spherically symmetric metric coefficients, by taking $X=X_{0}$, $X=X_{0}r^{m}$ and $A=A_{0}r^{m}, F=F_{0}r^{n}$ and in the sub-subsections $3.2.1$, $3.2.2$ and $3.2.3$ we find the Schwarzschild-de Sitter exterior solution, de Sitter solution and asymptotic solution, respectively. Finally we will finish with the conclusions.

\section{$\mathbf{f(T)}$ gravity theory} \label{sec2}

\subsection{Field equations}
To consider teleparallelism, one employs the orthonormal tetrad components
$e_A (x^{\mu})$, where an index $A$ runs over $0, 1, 2, 3$ to the
tangent space at each point $x^{\mu}$ of the manifold.
Their relation to the metric $g_{\mu\nu}$ is given by
\begin{equation}
g_{\mu\nu}=\eta_{A B} e^A_\mu e^B_\nu\,,
\label{eq:2.1}
\end{equation}
where $\mu$ and $\nu$ are coordinate indices on the manifold
and also run over $0, 1, 2, 3$,
and $e_A^\mu$ forms the tangent vector on the tangent space over which the
metric $\eta_{A B}$ is defined.

Instead of using the torsionless Levi-Civita connection in General Relativity,
we use the curvatureless Weitzenb\"{o}ck connection in teleparallelism
\cite{Weitzenb}, whose non-null torsion $T^\rho_{\verb| |\mu\nu}$ and contorsion
$K^{\rho}_{\verb| |\mu\nu}$ are defined by
\begin{eqnarray}
T^\rho_{\verb| |\mu\nu} \equiv \tilde{\Gamma}_{\nu\mu}^{\rho}-\tilde{\Gamma}_{\mu\nu}^{\rho}=e^\rho_A
\left( \partial_\mu e^A_\nu - \partial_\nu e^A_\mu \right)\,,
\label{eq:2.2}
\end{eqnarray}
\begin{eqnarray}
K_{\;\;\mu \nu }^{\rho} \equiv \tilde{\Gamma} _{\mu \nu }^{\rho}-\Gamma
{}_{\;\mu \nu }^{\rho}=\frac{1}{2}(T_{\mu }{}^{\rho }{}_{\nu }
+ T_{\nu}{}^{\rho }{}_{\mu }-T_{\;\;\mu \nu }^{\rho })\,
\label{eq:2.3}
\end{eqnarray}
respectively.
Here $\Gamma{}_{\;\;\mu \nu }^{\rho }$ is the
Levi-Civita connection.
Moreover, instead of the Ricci scalar $R$ for the Lagrangian density
in general relativity, the teleparallel Lagrangian density is described by the torsion scalar $T$ as follows:
\begin{equation}
T \equiv S_\rho^{\verb| |\mu\nu} T^\rho_{\verb| |\mu\nu}\,,
\label{eq:2.4}
\end{equation}
where
\begin{equation}
S_\rho^{\verb| |\mu\nu} \equiv \frac{1}{2}
\left(K^{\mu\nu}_{\verb|  |\rho}+\delta^\mu_\rho \
T^{\alpha \nu}_{\verb|  |\alpha}-\delta^\nu_\rho \
T^{\alpha \mu}_{\verb|  |\alpha}\right)\,.
\label{eq:2.5}
\end{equation}
%
The modified teleparallel action for $f(T)$ gravity
is given by~\cite{linder}
\begin{equation}\label{eq1}
{\cal S}=\int
d^{4}x|e|{f(T)}+\int d^{4}x|e|{\cal L_{M}},
\end{equation}
where $|e|= \det \left(e^A_\mu \right)=\sqrt{-g}$ and
the units have been chosen so that $c=16\pi G=1$.
Varying the action in equation~(\ref{eq1}) with respect to
the vierbein vector field $e_A^\mu$, we obtain the equation~\cite{rafael2}
\begin{equation}
\frac{1}{e} \partial_\mu \left( eS_A^{\verb| |\mu\nu} \right) F(T)
-e_A^\lambda T^\rho_{\verb| |\mu \lambda} S_\rho^{\verb| |\nu\mu}
F(T) +S_A^{\verb| |\mu\nu} \partial_\mu \left(T\right) F_{T}(T)
+\frac{1}{4} e_A^\nu f = \theta_A^\nu\,,
\label{eq:2.7}
\end{equation}
where a subscript $T$ denotes differentiation with respect to $T$ and $\theta_A^\nu$ is the matter energy-momentum tensor.

\subsection{Covariant field equations}
The field equation (\ref{eq:2.7}) is written in
terms of the tetrad and partial derivatives and to be appear
very different from Einstein's equations.
In this subsection, following \cite{barrow2}, we obtain an equation relating $T$ with
the Ricci scalar of the metric $R$. These will make the
equivalence between teleparallel gravity and general relativity
clear. On the other hand, the tetrad cannot be
eliminated completely in favor of the metric in equation (\ref{eq:2.7}),
because of the lack of local Lorentz symmetry, but we will
show that the latter can be brought in a form that closely
resembles Einstein's equation. This form is more suitable
for constructing spherical summitry solutions in the $f(T)$ theory.
To start writing the field equations in the covariant version,
we must replace  partial derivatives in the tensors by covariant derivatives compatible with the
metric $g_{\mu\nu}$, {\it i.e.} $\nabla_{\sigma}$  where $\nabla_{\sigma}g_{\mu\nu}=0$. Thus, equations (\ref{eq:2.2}), (\ref{eq:2.3}), and (\ref{eq:2.5}) can be written as
\begin{eqnarray}\label{1c}
T^{\rho}\,_{\mu\nu}=e^{\rho}_{A}(\nabla_{\mu}e^{A}_{\nu}-\nabla_{\nu}e^{A}_{\mu}),
\end{eqnarray}
where we have used the fact that $\Gamma^{\sigma}_{\mu\nu}$ is symmetric in the subscripts $\mu, \nu$:
\begin{eqnarray}\label{2c}
K^{\rho}\,_{\mu\nu}=e^{\rho}_{A}\nabla_{\nu}e^{A}_{\mu},
\end{eqnarray}
\begin{eqnarray}\label{3c}
S^{\mu\nu}\,_{\rho}=\eta^{AB}e^{\mu}_{A}\nabla_{\rho}e^{\,\nu}_{B}+\delta^{\nu}_{\rho}\eta^{AB}e^{\sigma}_{A}\nabla_{\sigma}e^{\,\mu}_{B}-
\delta^{\mu}_{\rho}\eta^{AB}e^{\sigma}_{A}\nabla_{\sigma}e^{\,\nu}_{B}
\end{eqnarray}
respectively.

On the other hand, from the relation between and
Weitzenb\"{o}ck connection and the Levi-Civita connection
given by equation (\ref{eq:2.3}), one can write the Riemann tensor for
the Levi-Civita connection in the form

\begin{eqnarray}\label{tensorR}
R^\rho_{\;\;\mu\lambda\nu}\!\!\!\!\!\!&&=\partial_{\lambda}\Gamma{}_{\;\mu\nu}^{\rho}
-\partial_{\nu}\Gamma {}_{\;\mu\lambda}^{\rho}
+\Gamma {}_{\;\sigma\lambda}^{\rho}\Gamma {}_{\;\mu\nu}^{\sigma}
-\Gamma {}_{\;\sigma\nu}^{\rho}\Gamma {}_{\;\mu\lambda}^{\sigma}\\ \nonumber
&&=\nabla_\nu K^\rho_{\;\;\mu\lambda}-\nabla_\lambda K^\rho_{\;\;\mu\nu}
+K^\rho_{\;\;\sigma\nu}K^\sigma_{\;\;\mu\lambda}-K^\rho_{\;\;\sigma\lambda}K^\sigma_{\;\;\mu\nu}\;,
\end{eqnarray}
whose associated Ricci tensor can then be written as
\begin{equation}
R_{\mu\nu}=\nabla_\nu K^\rho_{\;\;\mu\rho}-\nabla_\rho K^\rho_{\;\;\mu\nu}
+K^\rho_{\;\;\sigma\nu}K^\sigma_{\;\;\mu\rho}
-K^\rho_{\;\;\sigma\rho}K^\sigma_{\;\;\mu\nu}\;.
\end{equation}
Now, by using $K^\rho_{\;\;\mu\nu}$ given by equation~(\ref{eq:2.5}) along with the
relations $K^{(\mu\nu)\sigma}=T^{\mu(\nu\sigma)}=S^{\mu(\nu\sigma)}=0$ and considering that $S^\mu_{\;\;\rho\mu}=
2K^\mu_{\;\;\;\rho\mu}=-2T^\mu_{\;\;\;\rho\mu}$ one has
~\cite{barrow,barrow2,barrow3,reb}
\begin{eqnarray}
&&R_{\mu\nu}=-\nabla^\rho S_{\nu\rho\mu}-g_{\mu\nu}\nabla^\rho T^\sigma_{\;\;\;\rho\sigma}
-S^{\rho\sigma}_{\;\;\;\;\;\mu}K_{\sigma\rho\nu}\;, \nonumber \\
&&R=-T-2\nabla^\mu T^\nu_{\;\;\;\mu\nu}\;.
\end{eqnarray}
This last equation implies that the $T$ and $R$ differ only
by a covariant divergence of a space-time
vector. Therefore, the Einstein-Hilbert action and the teleparallel action
({\it i.e.} ${\cal S}=\int d^{4}x|e|T$) will both lead to the same field
equations and are dynamically equivalent theories. In Ref. \cite{barrow2} the authors have shown that
this equivalence is directly at the level of
the field equations. By using the equations listed
above and after some algebraic manipulations, one can gets

\begin{equation}\label{eqdivs}
G_{\mu\nu}-\frac{1}{2}\,g_{\mu\nu}\,T
=-\nabla^\rho S_{\nu\rho\mu}-S^{\sigma\rho}_{\;\;\;\;\mu}K_{\rho\sigma\nu}\;,
\end{equation}
where $G_{\mu\nu}=R_{\mu\nu}-(1/2)\,g_{\mu\nu}\,R$ is the Einstein tensor.

Finally, by using equation~(\ref{eqdivs}), the field equations for
$f(T)$ gravity equation~(\ref{eq:2.7}) can be rewritten in the form

\begin{equation}\label{eq2}
F(T)G_{\mu\nu}+\frac{1}{2}\left[f(T)-TF(T)\right]g_{\mu\nu}+B_{\mu\nu}F_{T}(T)=\theta_{\mu\nu},
\end{equation}
where $F(T)=\frac{df(T)}{dT}$ , $F_{T}(T)=\frac{dF(T)}{dT}$, $B_{\mu\nu}=S_{\nu\mu}\,^{\sigma}\nabla_{\sigma}T$ and $\theta_{\mu\nu}$ is the matter energy-momentum tensor.
Equation (\ref{eq2}) can be taken as the starting point of the $f(T )$
modified gravity model, and it has a structure similar to
the field equation of $f(R)$ gravity. Note that in the more general case with $f(T)\neq T$, the field equations
are covariant form. Nevertheless, the theory is not local Lorentz
invariant. In case of $f(T)=T$ and constant torsion, $f(T_0)$, GR is recovered and field equations are covariant
and the theory is Lorentz invariant.

\section{Spherically symmetric static solutions of $f(T)$ gravity}

In this section, we are looking for time-independent vacuum spherically symmetric solutions, henceforth the line element has the following form:

\begin{equation}\label{eq3}
ds^{2}=A(r)dt^{2}-B(r)dr^{2}-R^{2}d\Omega^{2},
\end{equation}
where $d\Omega^2 = dr^2+\sin^2\theta d\varphi^2$ and where $A$, $B$, and $R$ are three unknown functions.
One possible tetrad field (we can make arbitrary Lorentz transformations to the
tetrads without changing the metric) can be written as
\begin{equation}
e^{A}_{\mu}= {\rm diag}(A(r)^{1/2}, B(r)^{1/2}, R(r), R(r)\sin\theta) ,
\end{equation}
to which we refer to as the diagonal gauge.

In vacuum we have
\begin{equation}\label{eq4}
\theta_{\mu\nu}=0.
\end{equation}

Following \cite{vilja} under the assumption (\ref{eq4}), the new following form of equation (\ref{eq2}) (with fixed indices) is obviously index-independent:

\begin{equation}\label{eq5}
\frac{FR_{\mu\mu}+B_{\mu\mu}F_{T}}{g_{\mu\mu}}=K_{[\mu]}.
\end{equation}

 It is clear that for all $\mu$ and $\nu$ we can write $K_{[\mu]}-K_{[\nu]}=0$; therefore, having considered this result, the line element (\ref{eq3}) and equation (\ref{eq5}), we obtain two equations which include two unknown
parameters $A$ and $B$; thus the first equation arises by rewriting $K_{[t]}-K_{[r]}=0$;

\begin{equation}\label{eq6}
A''-\frac{2A}{r^{2}}-\frac{AX'}{rX}-\frac{FrX'}{2F_{T}}-\frac{A'X'}{X}=0,
\end{equation}
note that $X(r)=A(r)B(r)$ and also we have defined $'\equiv\frac{d}{dr}$.

The second equation will result from $K_{[t]}-K_{[\theta]}=0$;

\begin{eqnarray}\label{eq7}
8F_{T}XA^{2}+4F_{T}X'A^{2}r-4FAX^{2}r^{2}+2FXX'Ar^{3}-4F_{T}XAA'r+2F_{T}A'X'Ar^{2}-4r^{2}F_{T}XAA''\\\nonumber
+4Fr^{2}X^{3}-Fr^{4}XA'X'-2F_{T}r^{3}A'^{2}X'+2Fr^{4}X^{2}A''+2F_{T}A'r^{3}XA''=0.
\end{eqnarray}
The corresponding torsion scalar is

\begin{equation}\label{eq8}
T=2\frac{rA'+A}{r^{2}X},
\end{equation}
this equation describes that $T$ depends on $A(r)$ and $B(r)$, meanwhile since the metric only depends on $r$, one can say that $F(T)$ behaves like $F(r)$. Thus it will be certain that function $F(r)$ and it's derivative $F'(r)$ depend on the metric coefficients in a nonlinear and complicated way leading, to missing the analytical solutions of (\ref{eq6}) and (\ref{eq7}) in a closed form. The only way to deal with this situation is use of the numerical method.

 It should be noted that we are searching for physically acceptable solutions which includes analytical functions. So as to obtain analytical solutions for empty space in the $f(T)$ gravity model we are supposed to consider one or two assumptions about the form of metric coefficients ($X$ or $A$) or $F$ for each solution.

 As a first step in this direction, we consider $X=X_{0}$, which simplifies the equations.

\subsection{Solutions with $X=X_{0}$}
According to this assumption,  both equations (\ref{eq6}) and (\ref{eq7}) will be simplified. By unifying these equations, the corresponding metric coefficient will be given by

\begin{equation}\label{eq9}
A(r)=\frac{c_{2}}{r}+c_{1}r^{2},
\end{equation}
where $c_{1}$ and $c_{2}$ are integration constants. This equation directs us at $F(T)=F(T_{0})$ with constant torsion $T_{0}=\frac{6c_{1}}{X_{0}}$. This metric is a solution for any form of $f(T)$ for which there exists a constant $T_{0}$ such that $-T_{0}F(T_{0})+2f(T_{0})=0$, and $T_{0}$ should be real.

Because the dependence of $X$ on $r$, obviously we can also generalize the last constraint and for the next step we consider $X=X_{0}r^{m}$.

\subsection{Solutions with $X=X_{0}r^{m}$}
By solving the equations (\ref{eq6}) and (\ref{eq7}) for arbitrary $m$,  $A(r)$ and $F(r)$ are given by

\begin{equation}\label{eq10}
A(r)=\frac{2X_{0}r^{m}}{2+m-m^{2}}+c_{1}r^{2}+\frac{c_{2}}{r},
\end{equation}

\begin{equation}\label{eq11}
F(r)=F_{0}r^{m}.
\end{equation}
Also we easily can find $T$
\begin{equation}\label{eq12}
T=\frac{-4}{(m-2)r^{2}}+\frac{6c_{1}}{X_{0}r^{m}}.
\end{equation}
It has to be noted here that the equation $F=F_{0}r^{m}$ has a result exactly the same as the result of the assumption $X=X_{0}r^{m}$ and again $A$ and $T$ will be given by equations (\ref{eq10}) and (\ref{eq12}), and also equation (\ref{eq10}) shows that $m\neq2$.

To continue, we proceed to obtain  $f(T)$ by choosing $m=1$ to simplify the equations, specially the relation between $T$ and $r$. Thus for $T$ we have
\begin{equation}\label{eq13}
T=\frac{4}{r^{2}}+\frac{6c_{1}}{X_{0}r}.
\end{equation}
Having done a bit simplification, we lastly find $f(T)$:
\begin{equation}\label{eq144}
f(T)=\frac{F_{0}}{2} \left(\frac{-3c_{1}+\sqrt{4TX_{0}^{2}+9c_{1}^{2}}}{TX_{0}}\right)^{2},
\end{equation}
and assuming $\frac{9c_{1}^{2}}{4TX_{0}^{2}}<<1$, we expand $f(T)$ as follows:

\begin{equation}\label{eq155}
f(T)\simeq\frac{2F_{0}}{T}-\frac{6F_{0}c_{1}}{X_{0}}\left(\frac{1}{T^{3/2}}\right)+\left(\frac{9F_{0}c_{1}^{2}}{X_{0}^{2}}-
\frac{27F_{0}c_{1}^{3}}{4X_{0}^{3}}\right)\frac{1}{T^{2}}.
\end{equation}
Equation (\ref{eq144}) is a square quantity if $F_{0}>0$, thus we can say that in equation (\ref{eq155}) the positive terms are more effective than negative terms.
\subsubsection{Schwarzschild–-de Sitter solution}

Comparing the solution of equation (\ref{eq10}), to the well-known Schwarzschild--de Sitter metric, we have to set $m=0$, $X_{0}=1$ and require $c_{1}=-\frac{\Lambda}{3}$, $c_{2}=-2M$, therefore we can directly find $A(r)$ and $f(r)$:
\begin{equation}\label{eq13}
A(r)=1-\frac{2M}{r}-\frac{\Lambda}{3}r^{2},
\end{equation}
here $M$ represents the total mass which is result of a part of gravitational energy inside the sphere radius $r_{m}$, while the total energy also includes vacuum energy caused by a positive cosmological
constant. We have
\begin{equation}\label{eq14}
F(r)=F_{0},
\end{equation}
hence
\begin{equation}\label{eq15}
f(T)=F_{0}T+\lambda,
\end{equation}
where $\lambda$ is the integration constant.

\subsubsection{de Sitter solution}
The only difference between this solution and Schwarzschild–de Sitter solution is because of the value of $c_{2}$. Here if we assume $m=0$ and $c_{2}=0$ then we will have
\begin{eqnarray}\label{eq20}
A(r)=1-\frac{\Lambda}{3}r^{2}.
\end{eqnarray}
Note that the values of $f(T)$ and $T$ scalar have no change and they are thoroughly the same as their values in Schwarzschild–de Sitter solution.

  As you can see in equations (\ref{eq13}) and (\ref{eq20}) the Schwarzschild-de Sitter Solution; survives in the range $(r\rightarrow0)$ and the de Sitter solution survives in the range $(r\rightarrow\infty)$.
\subsubsection{Asymptotic solution}
Finding asymptotic solution needs the matching equation (\ref{eq10}) with the asymptotically flat solution, for which $A(r)\rightarrow1$ at large $r$. Due to this condition we should fix $m=0$, $c_{1}=0$, $c_{2}=-2M$, and $X_{0}=1$, so finally we gain the asymptotic metric coefficient

\begin{equation}\label{eq21}
A(r)=1-\frac{2M}{r},
\end{equation}
\begin{equation}\label{eq25}
T=\frac{2}{r^{2}}.
\end{equation}
It obvious that we have not gained any new results that tend to
constant scalar torsion ($T\rightarrow0$) in the large $r$ limit along with
$A(r)$, showing that any new solutions will be thoroughly
different from the Schwarzschild (-de Sitter) solution.

\subsection{Solutions with $A=A_{0}r^{m}, F=F_{0}r^{n}$}
By means of this assumption and simplifying equations (\ref{eq6}) and (\ref{eq7}), the unknown quantity $X(r)$ is given by

\begin{equation}\label{eq21}
X(r)=\frac{-A_{0}(m-2)(m+n+2)}{4r^{m+n+2}-A_{0}(m-2)(m+n+2)c_{1}}r^{2m+n+2}.
\end{equation}

By taking $m=-1$, $A$ and $X$ are given by
\begin{equation}\label{eq22}
A(r)=A_{0}/r.
\end{equation}
\begin{equation}\label{eq23}
X(r)=\frac{3A_{0}(n+1)r^{n}}{4r^{n+1}+3A_{0}(n+1)c_{1}}.
\end{equation}

Surprisingly, by substituting the above equations in the (\ref{eq8}) the torsion scalar vanishes and we cannot get any physically acceptable solutions. One can try these results for other values of $m$.

\section{Conclusions}
In this article we have obtained explicit solutions in a spherically symmetric space-time in $f(T)$ theory by choosing a diagonal form of the tetrad associated to the spherically symmetric metric. We started with modified field equations of $f(T)$ and rearranged them to have a class of equations involving some terms but not all terms of the field equations. Thereafter We went through just three special cases and determined the relations of metric coefficients of $F$. As a result we became able to reduce our unknown parameters and gain the metric coefficients. We extended our discussion and compared our results with the well-known Schwarzschild-de Sitter and de Sitter solutions and considered the asymptotic solution as well.

Taking notice of tetrad rules, the existence of tetrad in the $f(T)$ field equations causes this theory
not to respect local Lorentz transformation symmetry. Thus, a different field equations which in turn might have different solutions.
Some of these solutions do not have a valid GR counterpart, while others tend
to their GR counterparts in the appropriate limit. Therefore, special attention
has to be given to the choice of tetrad.
In Ref. \cite{bohmer} the authors have shown that there are two tetrad groups, bad tetrads and good tetrads. A good tetrad is the one that gives rise to field equations which do not constrain the functional
form of $f(T )$. In such cases one can always consider the limit $f(T )\rightarrow T$ where
the correct general relativistic limit is recovered. Otherwise we will talk of a
bad tetrad. We have studied spherical symmetric solutions in $f(T)$ theory by means of a diagonal tetrad, which are compatible with GR results; for an example refer to equation (\ref{eq13}).

,

\end{document}